\documentclass[bibyear]{aa}  
\usepackage{txfonts}
\usepackage{url}
\usepackage{enumitem}
\usepackage{appendix}
\usepackage{xcolor}
\usepackage{graphicx}
\usepackage{soul}

\def\spose#1{\hbox to 0pt{#1\hss}}
\newcommand\lsim{\mathrel{\spose{\lower 3pt\hbox{$\mathchar"218$}}
     \raise 2.0pt\hbox{$\mathchar"13C$}}}
\newcommand\gsim{\mathrel{\spose{\lower 3pt\hbox{$\mathchar"218$}}
     \raise 2.0pt\hbox{$\mathchar"13E$}}}

\def\ltsima{$\; \buildrel < \over \sim \;$}
\def\lsim{\lower.5ex\hbox{\ltsima}}
\def\gtsima{$\; \buildrel > \over \sim \;$}
\def\gsim{\lower.5ex\hbox{\gtsima}}

\def\apj{ApJ}
\def\apjl{ApJL}
\def\apjs{ApJSS}

\def\mnras{MNRAS}

\def\aap{A\&A}

\usepackage[normalem]{ulem}

\begin{document}

\title{The luminosity--volume test for cosmological Fast Radio Bursts}
\titlerunning{$V/V_{\rm max}$ of FRB}
\authorrunning{N. Locatelli, M. Ronchi, G. Ghirlanda, G. Ghisellini}
\author{N. Locatelli\inst{1,2}\thanks{E--mail: locatelli@ira.inaf.it}, M. Ronchi\inst{3,4}, G. Ghirlanda\inst{4,3,5}, G. Ghisellini\inst{4}  \\ }
\institute{$^1$  Università di Bologna, Dip. di Fisica \& Astronomia DIFA, 
    via Gobetti 92, Bologna, Italy \\
$^2$ INAF -- Istituto di Radioastronomia, via Gobetti 92, Bologna, Italy \\
$^3$ Università degli Studi di Milano-Bicocca, Dip. di Fisica "G. Occhialini", Piazza della Scienza 3, 20126, Milano, Italy\\
$^4$  INAF -- Osservatorio Astronomico di Brera, Via Bianchi 46, I--23807 Merate  Italy\\
$^5$  INFN - Sezione di Milano Bicocca, Piazza della Scienza 3, 20126, Milano, Italy 
   }


\abstract{We  apply the luminosity--volume test, also known as 
$\langle V/V_{\rm max}\rangle$, to Fast Radio Bursts (FRBs). We compare the 23 FRBs, recently discovered by ASKAP, with 20 of the FRBs found by Parkes. These samples have different flux limits and correspond to different explored volumes. 
We put constrains on their redshifts with probability distributions (PDFs) 
and apply the appropriate cosmological corrections to the spectrum and rate in order to compute the $\langle V/V_{\rm max}\rangle$ 
for the ASKAP and Parkes samples.
For a radio spectrum of FRBs ${\cal F}_\nu \propto \nu^{-1.6}$, we find 
$\langle V/V_{\rm max}\rangle=0.68\pm 0.05$ for the ASKAP sample, that includes FRBs up to $z=0.72_{-0.26}^{+0.42}$, and $0.54\pm 0.04$ for Parkes, that extends up to $z=2.1_{-0.38}^{+0.47}$.
The ASKAP value suggests that the population of FRB progenitors evolves faster 
than the star formation rate, while the Parkes value is consistent with it. Even a delayed (as a power law or gaussian) star formation rate cannot reproduce the $\langle V/V_{\rm max}\rangle$ of both samples.
If FRBs do not evolve in luminosity, the $\langle V/V_{\rm max}\rangle$ values of ASKAP and Parkes sample are consistent with a population of progenitors whose density strongly evolves with redshift as $\sim z^{2.8}$ up to $z \sim 0.7$. 
}
\maketitle

\section{Introduction}

Fast Radio Bursts (FRB) are very rapid ($\sim$0.1--10 ms) and bright ($\sim$0.1--1 Jy) radio pulses  typically observed at $\sim$1 GHz frequency.
Their detection is characterized by a frequency--dependent delay ($\propto \nu^{-2}$) of the arrival time and a frequency--dependent broadening ($\propto \nu^{-4}$) of the radio signal. These are typical signatures of a signal propagating  through a low-density relativistic plasma.
The signal dispersion is quantified by the Dispersion Measure (DM), proportional to the integral of the free electron density $n_e$ along the line-of-sight (LOS) from the observer to the source.

A collection of 52 FRBs has been made public through a database\footnote{Fast Radio Burst Catalog (FRBCAT)  {http://frbcat.org/}} by Petroff et al. (2016)$^{\cite{petroff16}}$. 
The large values of the observed dispersion measure ${\rm DM}_{\rm obs}$ (Lorimer et al., 2007$^{\cite{lorimer07}}$; Thornton et al., 2013$^{\cite{thornton13}}$) and the 
detection of the host galaxy for the repeating FRB 121102 (Marcote et al., 2017\cite{marcote17}), seem to favour their extra--galactic nature. 
The host galaxy redshift $z \sim 0.19$ (Chattarjee et al., 2017$^{\cite{chattarjee17}}$, Tendulkar et al., 2017$^{\cite{tendulkar17}}$)  is in fact consistent with the distance estimated from the observed DM. 
Despite the confirmed extra--galactic origin of this source, the
fact that it remains the only repeater holds the possibility that it could be representative of a different class. 

The ``debate" on the origin of FRBs reminds much of the similar case of Gamma Ray Bursts (GRBs) (see Kulkarni 2018\cite{kulkarni18}). 
Indeed, if  extragalactic, FRBs have $\nu L_\nu$ luminosities 
around $10^{43}$ erg s$^{-1}$ and energetics of the order of 
$10^{40}$ erg (Fig.~\ref{fig:E,z_distribution}).

\begin{figure*}
\centering
\includegraphics[width=\textwidth]{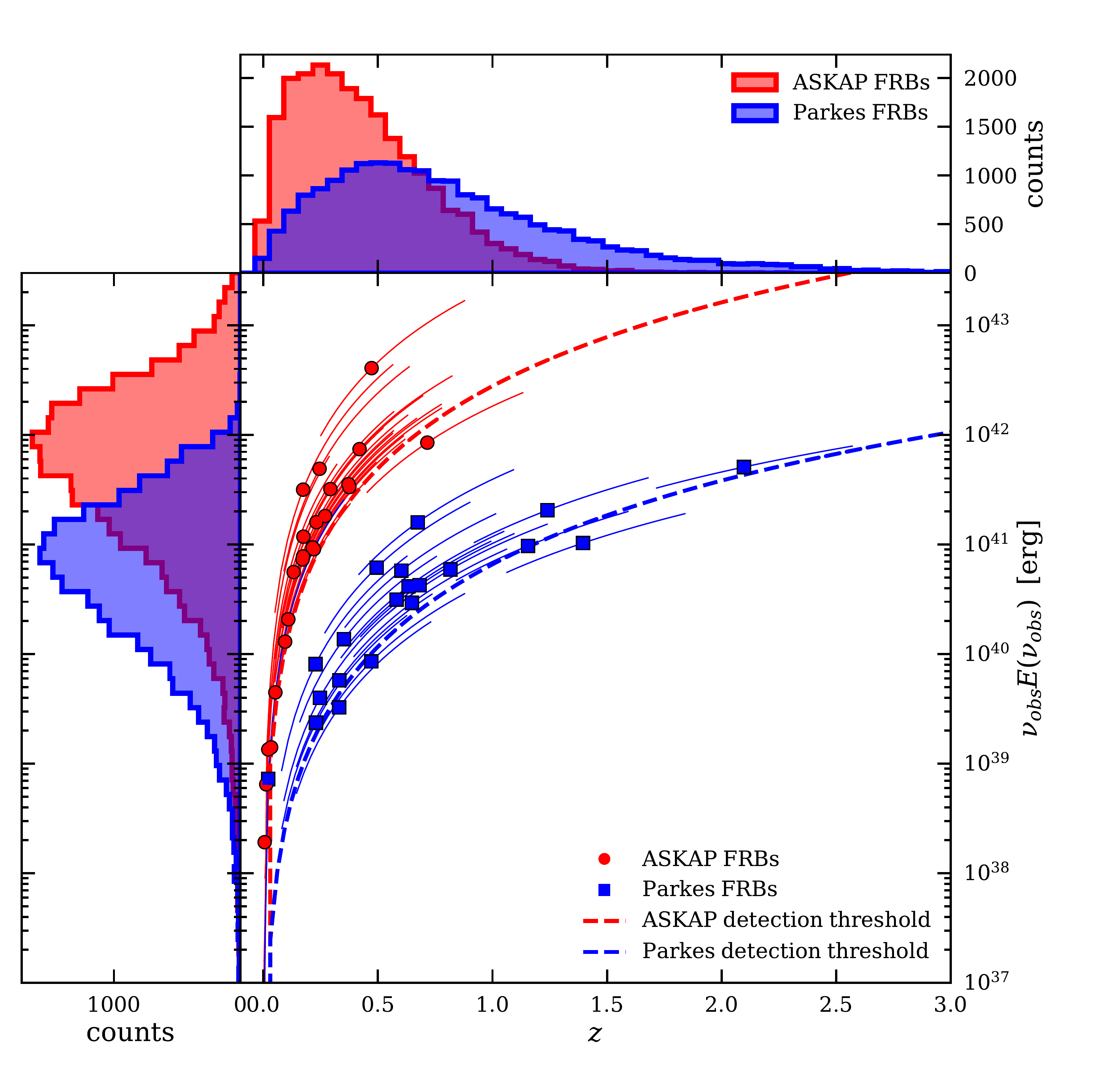}
\caption{Energy ($\nu E_\nu$) as a function of redshift for FRBs 
detected by ASKAP (red circles) and Parkes (blue squares). 
The dashed lines correspond to the fluence limit for ASKAP and Parkes (red and blue respectively) obtained assuming the average FRB duration of the respective samples. The top and right panels show the energy and redshift distributions of the two samples (same colour coding of the central plot). The thin curves  represent, for each data point, the 1$\sigma$ uncertainty on the energy produced by the  uncertainty  on the redshift. }
\label{fig:E,z_distribution}
\end{figure*}
Such large energies and short durations imply a huge brightness temperature, of the order of $T_{\rm B}\sim 10^{34}-10^{37}$ K.
This in turn requires a coherent radiation process possibly originating from 
masers or compact bunches of emitting particles, as it has been recognized by many authors (Ghisellini 2017$^{\cite{ghisellini17}}$; Kumar, Lu \& Bhattacharya 2017$^{\cite{kumar17}}$; Yang \& Zhang 2017$^{\cite{yang17}}$; Ghisellini \& Locatelli 2018$^{\cite{ghisellini18}}$; Katz 2018$^{\cite{katz18}}$).

Many of the proposed progenitor theories of extragalactic FRBs include merging of compact objects such as neutron stars (Totani, 2013$^{\cite{totani13}}$); 
or white dwarfs (Kashiyama et al., 2013$^{\cite{kashiyama13}}$).
FRBs could be flares from magnetars 
(Popov \& Postnov 2010$^{\cite{totani13}}$; Thornton et al., 2013$^{\cite{thornton13}}$; 
Lyubarsky 2014$^{\cite{lyubarsky14}}$; 
Beloborodov 2017$^{\cite{belobodorov17}}$) or 
giant pulses from pulsars 
(Cordes \& Wasserman 2015$^{\cite{cordes15}}$), 
or they could be associated to  
the collapse of supra-massive neutron stars 
(Zhang 2014; Falcke \& Rezzolla (2014)$^{\cite{falcke14}}$);
or dark matter induced collapse of neutron stars 
(Fuller \& Ott 2015$^{\cite{fuller15}}$). 
See Platts et al. 2018\cite{platts18}, for an updated 
list of FRB theories. 

A classical way to probe the unknown distance of a population of objects is through the $\langle V/V_{\rm max}\rangle$ test (also called \textit{luminosity--volume} test). 
Firstly proposed by Schmidt (1968)$^{\cite{schmidt68}}$, the
$\langle V/V_{\rm max} \rangle$ tests whether the distribution of objects is uniform within the volume of space defined by the observational selection criteria. Among other advantages, it is  suitable for samples containing few objects and allows to combine samples of sources obtained with different selection criteria. Historically, it has been employed to study the space distribution of quasars and to assess the cosmic evolution of their population.

For a uniform population of sources with measured fluxes $S$,  $V/V_{\rm max}$ are the ratios of the volume $V$  within which each source is distributed to the maximum volume  $V_{\rm max}$ within which each source could still be detected (which is individually defined by the sample selection flux limit). In an Euclidean space $V/V_{\rm max}$  should be uniformly distributed between 0 and 1 with an average value $\left\langle V/V_{\rm max} \right\rangle=0.5$. 
Equivalently the cumulative source count distribution is  $N(>S)\propto S^{-3/2}$. 

The standard Euclidean approach to the luminosity--volume test has been performed  on FRBs by different authors: 
Oppermann et al. (2016) \cite{oppermann16} find $N(>S)\propto S^{-n}$ with 
$0.8\leq n\leq1.7$; 
Caleb et al. (2016)\cite{caleb16}, report $n=0.9\pm0.3$ while 
Li et al. (2016)\cite{li16} find a much flatter population with $n=0.14\pm0.20$. More recently, 
James et al.  (2018)\cite{james18} (hereafter J18), updated these results with 23 new FRBs (Shannon et al., 2018\cite{shannon18}) discovered by the ASKAP array through the CRAFT survey (Macquart et al., 2010\cite{macquart10}). They find
$n=1.52\pm0.24$ for the combined ASKAP and Parkes FRB samples. 
However, for the first time, they compare different surveys with sufficient statistics, claiming a significant difference between the ASKAP CRAFT ($n=2.20\pm0.47$) and Parkes HTRU ($n=1.18\pm0.24$) surveys, respectively. 

However, the assumption of a uniform distribution in Euclidean space for the 
number counts -- $\langle V/V_{\rm max} \rangle$ test does not deal with the transient nature of FRBs. 
In this paper we perform the volume--luminosity test using the redshift estimated through the DM. We account for the possible uncertainty on $z$ in terms of a probability density function (PDF) which encodes  
the uncertainties on i) the Galactic free electron density model; ii) the baryon distribution in the Inter-Galactic Medium (IGM) and iii) the free electron density model of the FRB host environment.
Furhtermore, we apply to each FRB the appropriate K--correction
and account for the proper transformation from the observed to the intrinsic rate.

In \S\ref{sec:samples} we present the two FRB samples used; 
in \S\ref{sec:methods} we describe how to 
derive their $\langle V/V_{\rm max}\rangle$ accounting for their cosmological nature. 
A Monte Carlo approach is adopted to perform the  $\langle V/V_{\rm max}\rangle$ 
for cosmological rate distributions (\ref{subsec:sim}) and compare with the real 
samples in \S\ref{sec:results}. 
In \S 5 we discuss our results, and in \S 6 we draw our conclusions. We adopt a standard cosmology with $\Omega_{\rm max}=0.286,\, h=0.696$ and $\Omega_{\Lambda}=0.714$. 

\section{Data samples}\label{sec:samples}

The ASKAP array of radio telescopes and the Parkes telescope provided two relatively large and well defined samples of FRBs which allow us to perform statistical analysis.
From the online catalogue FRBCAT\footnote{available at http://www.frbcat.org} (Petroff et al. 2016), we collected all ASKAP and Parkes FRBs that have been confirmed via publication with a known signal to noise ratio (S/N) threshold.

\subsection{ASKAP}\label{subsec:askap}

We consider the 23 FRBs detected by ASKAP and recently published by Shannon et al. (2018)\cite{shannon18} and  Macquart et al. (2018)\cite{macquart18}. The ASKAP survey has an exposure of $5.1 \times 10^5 \ \rm deg ^2 \ \rm h$, a field of view of $20 \ \rm deg^2$ and a unique S/N threshold of $9.5$ which corresponds to a limiting fluence of  $23.16\times (w_{\rm obs}/1 \ \rm ms)^{1/2} \ \rm Jy \ \rm ms \ $ (Shannon et al. 2018). 
Although FRB 171216 has been detected with a $S/N=8$ in a single beam, it has a $S/N=10.3$ considering
its detection in the two adjacent beams (Shannon et al. 2018) and we then include this event in our sample.

The ASKAP sample is shown in Fig. \ref{fig:E,z_distribution} (red circles). To properly compare the energy density of FRBs at different redshifts we evaluate the energies at the observed frequency $\nu_{\rm obs} = 1.3$ GHz for all the FRBs. 
We apply the K--correction assuming an energy power law spectrum $E_{\nu} \propto \nu^{-\alpha}$ with $\alpha = 1.6$ (Macquart et al. 2018). 
Therefore the K--corrected energy density is given by Eq. \ref{eq:lum_rel} (see \appendixname{ A}) 
$E_{\nu}(\nu_{\rm obs}) = E_{\nu}(\nu_{\rm rest}) (1+z)^{\alpha}$. 
The red dashed line corresponds to the ASKAP limiting fluence assuming the average intrinsic duration of the ASKAP sample $\langle w_{\rm rest} \rangle=2.2$~ms. 
The observed pulse duration (which defines the limiting curve - see above) scales as $w_{\rm obs} = \langle w_{\rm rest} \rangle (1+z)$. 
Because of the large field of view  and the high fluence threshold, the ASKAP survey is more sensitive to nearer and relatively powerful events. 
This is evident in the redshift and energy distributions (top and left panels of Fig.~\ref{fig:E,z_distribution}) of the ASKAP sample whose mean values are $\langle z\rangle = 0.25$ and $\langle E\rangle = 4.3 \times 10^{41} \ \rm erg$ respectively.

\subsection{Parkes}\label{subsec:parkes}

Among all Parkes FRBs we found 20 verified events with known S/N threshold. 
In this case, since FRBs were detected in various surveys with different instrumental setups, the S/N threshold is not equal for all FRBs and therefore the corresponding fluence limit is not unique. 
The mean fluence limit of this sample results $0.54\times(w_{\rm obs}/1 \ \rm ms )^{1/2} \ \rm Jy \ \rm ms \ $ and is represented by the dashed blue line in Fig.~\ref{fig:E,z_distribution}. 
Compared with ASKAP, the Parkes telescope is characterised by a smaller field of view ($\sim 0.01 \,{\rm deg}^2$) and a lower fluence limit being thus  
sensitive to more distant and less powerful events on average (cf. the distributions in the top and left panels of  Fig.~\ref{fig:E,z_distribution}). The total exposure of the HTRU survey made at Parkes is $1441 \ \rm deg ^2 \ \rm h$ (Champion et al., 2016\cite{champion16}).
The mean redshift and energy of the Parkes sample are $\langle z\rangle = 0.67$ and $\langle E\rangle = 7.2 \times 10^{40} \ \rm erg$, respectively.

\vskip 0.5 cm

The total sample includes 43 FRBs. When different data analyses for the same FRB were found, we chose the one computed with the method presented in Petroff et al. (2016)$^{\cite{petroff16}}$ 
(if available) in order to build a uniform sample and because alternative searches tend to under--estimate the $S/N$ (see Keane \& Petroff 2015$^{\cite{keane15}}$). 
Table \ref{tab:sample_obs} lists the galactic longitude and 
latitude ($gl$ and $gb$), the observed dispersion measure 
($DM_{\rm obs}$), the signal to noise ratio ($S/N$), 
the survey S/N threshold ($S/N_{\rm lim}$), fluence ${\cal F}(\nu_{\rm obs})$, duration $w_{\rm obs}$ and flux density $S(\nu_{\rm obs})$ all evaluated at the observed frequency $\nu_{\rm obs}$, the redshift calculated as presented in $\S 3$ and the references of the S/N$_{\rm lim}$ for all the 43 FRBs in our sample.

\section{Method}\label{sec:methods}

Redshift information is necessary to perform the luminosity-volume test for a cosmological population of sources.
In order to estimate $z$, Ioka et al. (2003)\cite{ioka03} and Inoue et al. (2004)\cite{inoue04} propose a linear relation between the redshift and the dispersion measure due to the IGM (${\rm  DM}_{\rm IGM}$):
\begin{equation}
{\rm DM}_{\rm IGM} = C \cdot z \label{eq:DM_to_z}
\end{equation}
where $C = 1200 \, [\mbox{pc/cm}^3]$. The redshift of a transient radio source can be estimated from the  residual dispersion measure DM$_{\rm excess}$, i.e. once the Milky Way contribution is subtracted from the observed DM$_{\rm obs}$ measured for the particular source:
\begin{eqnarray} \label{eq:z_peak}
    {\rm DM}_{\rm excess} &=& {\rm DM}_{\rm obs}-{\rm DM}_{\rm MW} \nonumber \\
    &=&   \frac{{\rm DM}_{\rm host}}{1+z} +{\rm DM}_{\rm IGM} 
\end{eqnarray}
where DM$_{\rm host}$ is the dispersion measure of the host galaxy environment. 
The Milky Way dispersion measure DM$_{\rm MW}$ is estimated in this work using the model of Yao, Manchester \& Wang (2016)\cite{yao17} (YMW16 hereafter).
The redshift of the source can be found from Eq. \ref{eq:z_peak}.

In principle, a large uncertainty on the redshift estimate is expected due to either the local environment and host galaxy free electron density and inclination (Xu \& Han, 2015\cite{xu15}, Luo et al., 2018\cite{luo18}) and to the high variance $\sigma^2_{\rm DM}(z)$ of the unknown baryon halos and sub-structures along the LOS (McQuinn 2014\cite{mcquinn14}, Dolag et al. 2016\cite{dolag15}).
In order to take into account these two sources of uncertainty -- affecting DM$_{\rm host}$ and DM$_{\rm IGM}$  respectively in Eq.~\ref{eq:z_peak} -- we calculate the $\langle V/V_{\rm max}\rangle$ building appropriate probability functions for the redshifts PDF($z$), rather than a unique value as obtained through Eq.~\ref{eq:DM_to_z}.

Firstly we need to model the host galaxy contribution DM$_{\rm host}$.
Xu \& Han (2015)\cite{xu15} estimate DM$_{\rm host}$ for different galaxy morphologies and inclination angles $i$ with respect to the LOS. 
In particular, for spiral galaxies they fit skew normal functions $f(\rm DM,i)$ to the DM$_{\rm host}$ distributions for different values of the inclination angle $i$. 
We find the overall ${\rm DM}_{\rm host}$ PDF ($P({\rm DM}_{\rm host})$) by averaging the functions $f(\rm DM,i)$, each weighted with the probability of the inclination angle $\sin i$. We extract the DM$_{\rm host}$ values from $P({\rm DM}_{\rm host})$.
This allows us to calculate the redshift $z_{\rm peak}$, corresponding to the most probable value to be associated to the ${\rm DM}_{\rm excess}$ of the $j$--th FRB.
The DM distribution of the host galaxy and environment has been also estimated by Luo et al. (2018). 
They obtain in general a smaller contribution of ${\rm DM}_{\rm host}$ to the ${\rm DM}_{\rm excess}$ than what derived by Xu \& Han.(2015). 
This is probably due to the simplistic but more conservative assumption of a MW, M31-like galaxy as a host environment in the latter work, from which we derive the ${\rm DM}_{\rm host}$ distribution.\\

The other source of uncertainty is related to the variance of ${\rm DM}_{\rm IGM}$. 
This has been studied by e.g. McQuinn (2014) and Dolag et al. (2015).
We considered the scenario giving the largest $\sigma_{\rm DM}(z)$ as reported in McQuinn (2014), obtained with a baryon (i.e. $n_e$) distribution tracing the dark matter halos above a certain mass threshold. 
Although dedicated simulations show smaller uncertainties (McQuinn 2014, Dolag et al. 2016) we choose the most conservative $\sigma_{\rm DM}(z)$ relation, represented by the power-law function:
\begin{equation}
    \sigma_{\rm DM}(z) = \frac{379.2 }{C_{1200}} \, z^{0.313}
    \label{eq:fit_sigmaDM}
\end{equation}
where $C_{1200}$ is the coefficient $C$ in units of $1200 {\rm pc \,cm}^{-3}$.

We calculate an inferior ($z_{\rm inf}$) and superior value ($z_{\rm sup}$) for the redshift in the PDF 
by introducing the left and right standard deviations $\sigma_{\rm DM}(z)$ respectively in the RHS of Eq.~\ref{eq:z_peak}:

\begin{eqnarray}
    DM_{\rm excess} \equiv & C\cdot z_{\rm inf} + \sigma_{\rm DM}(z_{\rm inf}) +\frac{{\rm DM}_{\rm host}}{1+z_{\rm inf}}\\ 
    DM_{\rm excess} \equiv & C\cdot z_{\rm sup} - \sigma_{\rm DM}(z_{\rm sup}).
    \label{eq:z_inf_sup}
\end{eqnarray}

In the second equation a null ${\rm DM}_{\rm host}$ contribution gives an upper limit to the redshift distribution.

The PDF is then shaped as an asymmetric Gaussian: the peak of the PDF is assigned to the redshift $z_{\rm peak}$; the left and right dispersion are $\sigma_{\rm DM}(z_{\rm inf})=z_{\rm peak}-z_{\rm inf}$ and $\sigma_{\rm DM}=z_{\rm sup}-z_{\rm peak}$ respectively, where $z_{\rm inf}<z_{\rm peak}<z_{\rm sup}$.\\

To calculate the Galactic contribution to the DM (DM$_{\rm MW}$) we considered two $n_e$ models: 
the NE2001 model (Cordes~\& Lazio, 2001$^{\cite{cordes02}}$) and 
the model of Yao, Manchester \& Wang (2016)\cite{yao17} (YMW16)
\footnote{Online calculators are available respectively at: \\
https://www.nrl.navy.mil/rsd/RORF/ne2001/model.cgi\\
http://www.atnf.csiro.au/research/pulsar/ymw16}. 
Assuming different models for the MW free electron contribution does not significantly affect the redshift distribution of the FRB sample. 
The same result has also been found by Luo et al. (2018) who also consider the presence of a free electron dark halo around the Milky Way. 
Its effect is found to be negligible however.
This enables us to base our analysis assuming one of the DM models without 
loosing generality. 
We consider the YMW16 as our baseline model.\\

Up to now we accounted for the stochastic dispersion around the DM$_{\rm IGM}(z)$ relation.
However, an additional source of uncertainty can systematically arise from the choice of the average DM$_{\rm IGM}(z)$ relation, namely the $C$ coefficient of Eq.~2. 
In fact simulations (Dolag et al., 2015\cite{dolag15}) show a lower value than the one derived from modelling of the free electron density in the IGM. 
To account also for this systematic effect we calculate redshift uncertainties assuming different values of $C$ in the range $[950,1200]$.

In general, large values of $C$ decrease both the redshift $z$ and the maximum observable redshift $z_{\rm max}$, but by a slightly different amount, making the average $\langle V/V_{\rm max}\rangle$ to also decrease.
However, the relative change of the $\langle V/V_{\rm max}\rangle$ values, for the $C$ values considered, is limited to less than 3$\%$ for all spectral indexes.
For semplicity, we thus assume $C=1200$.

\subsection{$\langle V/V_{\rm max}\rangle$ of the FRB samples}

The cosmological $\langle V/V_{\rm max}\rangle$ test is defined as the average over the sample of the ratios between the volume of space included within the source distance and the maximum volume in which an event, holding the same intrinsic properties, could have been observed. In terms of comoving distance $D(z)$: 
\begin{equation}
\left\langle \frac{V}{V_{\rm max}} \right\rangle = \frac{1}{N}\sum_{i=0}^N \left[\frac{D^3(z)}{D_{\rm max}^3(z_{\rm max})}\right]_i
\label{eq:VVmax_1}
\end{equation}
where $N$ is the total number of FRBs in one sample, $D(z)$ and $z$ are, respectively, the comoving distance and redshift of each object. $D_{\rm max}$ and 
$z_{\rm max}$ are the same quantities that would be evaluated if that same source was observed at the limiting threshold of the same survey that found it, i.e. if its measured fluence ${\cal F}_{\nu}$ (or $S/N$) coincided with the survey detection thresholds, ${\cal F}_{\nu,lim}$ (or $S/N_{\rm lim}$), assuming the same intrinsic properties of the event.

A large value of $\langle V/V_{\rm max}\rangle$ indicates sources distributed closer to the boundary of the volume probed by the given survey. Conversely, a low $\langle V/V_{\rm max}\rangle$ value indicates a population  distributed closer to the observer with respect to the total survey--inspected volume.

The largest volume which can be probed $\propto D^3_{\rm max}(z_{\rm max})$ is uniquely defined by the survey/instrumental parameters once a cosmology is assumed. 
It can be expressed in terms of the maximum possible redshift $z_{\rm max}$ at which a given source can be observed  with the survey considered. 
In practice, we solve for $z_{\rm max}$ the equation (see \appendixname{ A} for the complete derivation):
\begin{equation}
  \frac{D(z)}{D_{\rm max}(z_{\rm max})} =  \left[ \frac{S/N_{\rm lim}}{S/N}  \left(\frac{1+z}{1+z_{\rm max}}\right)^{-\frac{1}{2}-\alpha}\right]^\frac{1}{2} \label{eq:D/Dm}
\end{equation}
for each FRB in our sample. 
Here $\alpha$ is the observed spectral index of FRBs. Our formulation of the luminosity--volume test has been implemented for transient events embedded in a non-Euclidean cosmological volume and can be used with any user-defined luminosity function, source distribution, source spectral index and cosmology.

Following Shannon et al. (2018), we assume a power--law spectrum for the whole FRB population, with non--evolving spectral index $\alpha$. 
We consider three possible values of the spectral index $\alpha = 0$, 
$\alpha = 1.6$ (see Macquart et al. 2018) and $\alpha = 3$ 
in order to test how the spectral slope affects the estimate of 
$\langle V/V_{\rm max}\rangle$. 
Redshift are obtained through a Monte-Carlo extraction, as described at the beginning of this section.
By estimating $z_{\rm max}$ for each FRB we can compute 
$\langle V/V_{\rm max}\rangle$ for the two samples of ASKAP and Parkes FRBs and for the full combined ASKAP+Parkes sample. 
We then repeat the test $10^4$ times to account for the stochastic uncertainty on redshift. 
Considering a spectral index of $1.6$ as found by Macquart et al. (2018), we find $\langle V/V_{\rm max}\rangle = 0.681 \pm 0.049$ and $\langle V/V_{\rm max}\rangle = 0.538 \pm 0.046$ for ASKAP and Parkes, respectively, and $\langle V/V_{\rm max}\rangle = 0.624 \pm 0.086$ for the full sample. 
Values of $\langle V/V_{\rm max}\rangle$ for different values of the spectral index $\alpha$ are reported in Table \ref{tab:V/Vm_obs}.
\begin{table}
\centering
\begin{tabular}{c|ccc}
$\alpha$ & ASKAP & Parkes & full sample\\
\hline
0    & $0.634\pm0.054$ & $0.443\pm0.048$ & $0.538\pm0.109$\\
1.6 & $0.681\pm0.049$ & $0.538\pm0.046$ & $0.624\pm0.086$\\
3   & $0.711\pm0.046$ & $0.593\pm0.045$ & $0.652\pm0.075$\\
\hline
$\langle z \rangle$ & $0.429 \pm 0.063$ & $0.820\pm 0.120$ & $0.624\pm0.218$
\end{tabular}
\vskip 0.2 cm
\caption{$\langle V/V_{\rm max}\rangle$ for different values of the spectral index $\alpha$. Values are obtained for the full sample and for the ASKAP and Parkes sub-samples. They assume $C= 1200$. The mean redshift $\langle z \rangle$ in each sample is also reported.
}
    \label{tab:V/Vm_obs}
\end{table}
\begin{table}
\centering
\begin{tabular}{l | l l l}
$\langle z \rangle$ &CSFR (MD14) & SGRBs (G16) &Const. \\
\hline
 0.429  &0.536 &0.519 &0.474 \\
 0.624 &0.544 &0.519 &0.464 \\
 0.820 &0.549 &0.516 &0.457 \\
\hline
    \end{tabular}
\vskip 0.2 cm
\caption{$\langle V/V_{\rm max}\rangle$ obtained at different $\langle z \rangle$ for populations extracted via the Monte Carlo method from the three tested redshift distributions described in the text. }
\label{tab:V/Vm_sim}
\end{table}

\subsection{Comparison with simulated populations}
\label{subsec:sim}

We compare the values of the $\langle V/V_{\rm max}\rangle$ obtained for the ASKAP and Parkes samples with the $\langle V/V_{\rm max}\rangle$ expected for different cosmological population of sources. We tested three different redshift density distributions:
\begin{itemize}
    \item[1)] the cosmic star formation rate (Madau \& Dickinson, 2014, hereafter MD14);
    \item[2)] the short GRBs redshift distribution (as found by Ghirlanda et al., 2016, hereafter G16);
    \item[3)] a constant density distribution (i.e. no evolution).
\end{itemize}

Let us call ``energy function" (EF) the density of sources (in the comoving volume) as
a function of their radiated energy (in analogy with the luminosity function).
For simplicity, we here neglect any EF evolution in cosmic time, considering only the density evolution corresponding to three cases above.

\begin{figure*}
\centering
\includegraphics[width=\textwidth]{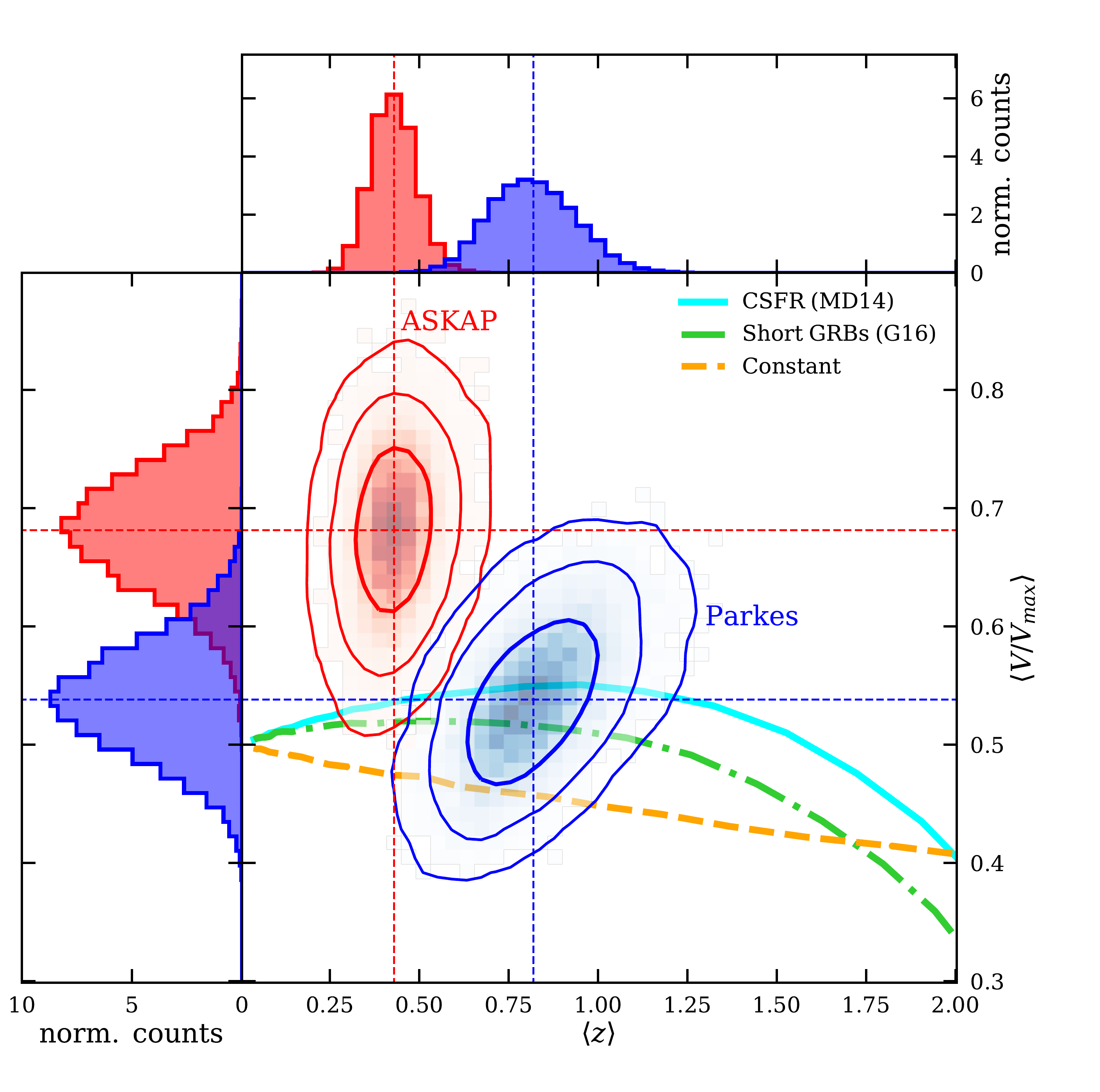}
\caption{ $\langle V/V_{\rm max}\rangle$ of the ASKAP and Parkes FRB samples (red and blue squares, respectively) computed accounting for the cosmological terms (K--correction and rate) and the redshifts dispersion. Contours represent the 1-, 2- and 3-$\sigma$ uncertainties on $\langle V/V_{\rm max}\rangle$. A spectral index   
$\alpha=1.6$ is assumed in this figure. Similar plots obtained with other possible values of the spectral index are shown in Fig.~\ref{fig:VVmVSzm_alfa}. 
The lines show the trends of $\langle V/V_{\rm max}\rangle$ as a function of increasing average redshift (i.e. survey depth). 
The different colours (styles) thick lines show the results of the evolution of $\langle V/V_{\rm max}\rangle$ obtained assuming different star formation rates (as labelled, see also \S \ref{sec:discussion}). }
\label{fig:VVmVSzm}
\end{figure*}

Synthetic populations were generated through a Monte Carlo extraction from a PDF proportional to a given $\psi_k(z)$ which represents the source density (i.e. per unit  comoving volume) rate (i.e. per unit comoving time). 
The subscripts $k=1,2,3$ refer to the three cases considered above. 
By accounting for the cosmological time dilation and volume, the sampling probability density function is: 
\begin{equation}
	\mbox{pdf}_k(z)\,dz \propto \psi_k(z)\cdot(1+z)^{-1} \frac{dV}{dz}(z)\, dz \label{eq:pdf_def}
\end{equation}
We then use the following procedure:
\begin{enumerate}
    \item 
We assume a value of $z_{\rm max}$ as the maximum redshift at which a FRB population can be detected and generate a fake FRB sample from $z = 0$ to $z = z_{\rm max}$.

\item
Then we evaluate the average $\langle V/V_{\rm max}\rangle$ and 
average $\langle z \rangle$ of the synthetic sample corresponding to that $z_{\rm max}$. 

\item
The $\left\langle V/V_{\rm max} \right\rangle$ calculated from these events is then assigned to the mean redshift $\langle z \rangle$ of the simulated FRBs. 
It becomes a point of the plotted curve in Fig. \ref{fig:VVmVSzm}.

\item
We repeat this procedure for all values of  $z_{\rm max}$ in order to obtain 
the model curves in Fig. \ref{fig:VVmVSzm}. 

\end{enumerate}

The value of $z_{\rm max}$ is linked to the value of the intrinsic energy of the FRB through Eq. \ref{eq:ups_def}, once the dependence of S/N on the energy, intrinsic duration and distance is made explicit 
(see \appendixname{ A}):
\begin{equation}
   D^2_{\rm max}(z_{\rm max}) (1+z_{\rm max})^{\alpha+\frac{1}{2}} = \frac{E_\nu(\nu_{\rm obs})}{4\pi A \, (w_{\rm rest}/ \rm ms)^{1/2}}  \label{eq:zM_sim_find}
\end{equation}
where $D_{\rm max}(z_{\rm max})$ is the proper distance at redshift $z_{\rm max}$; $E_\nu(\nu_{\rm obs})$ is the energy density of the FRB at the observed frequency $\nu_{\rm obs} \simeq 1.3 \; \rm GHz$.
The constants $A=23.16$ Jy ms for ASKAP and $A=0.54$ Jy ms for Parkes   specify the fluence limit of the two instruments: 
\begin{equation}
    F_{\rm lim}(\nu_{\rm obs}) = A \sqrt{\frac{w_{\rm obs}'}{\mbox{ms}}}\, = A \left(\frac{w_{\rm rest}}{\mbox{ms}}\right)^{1/2} (1+z_{\rm max})^{1/2}\,\, [{\rm Jy\, ms}] \label{eq:F_lim}
\end{equation}
here $w_{\rm obs}' = w_{\rm rest} (1+z_{\rm max})$ is the pulse duration we would 
observe if the FRB were located at redshift $z_{\rm max}$. 
Therefore fixing $z_{\rm max}$ for a given extraction implies choosing the same intrinsic energy density $E_{\nu}(\nu_{\rm obs})$ for all FRBs of the fake sample generated according to that $z_{\rm max}$.

\begin{figure*}
\centering
\includegraphics[width=0.49\textwidth]{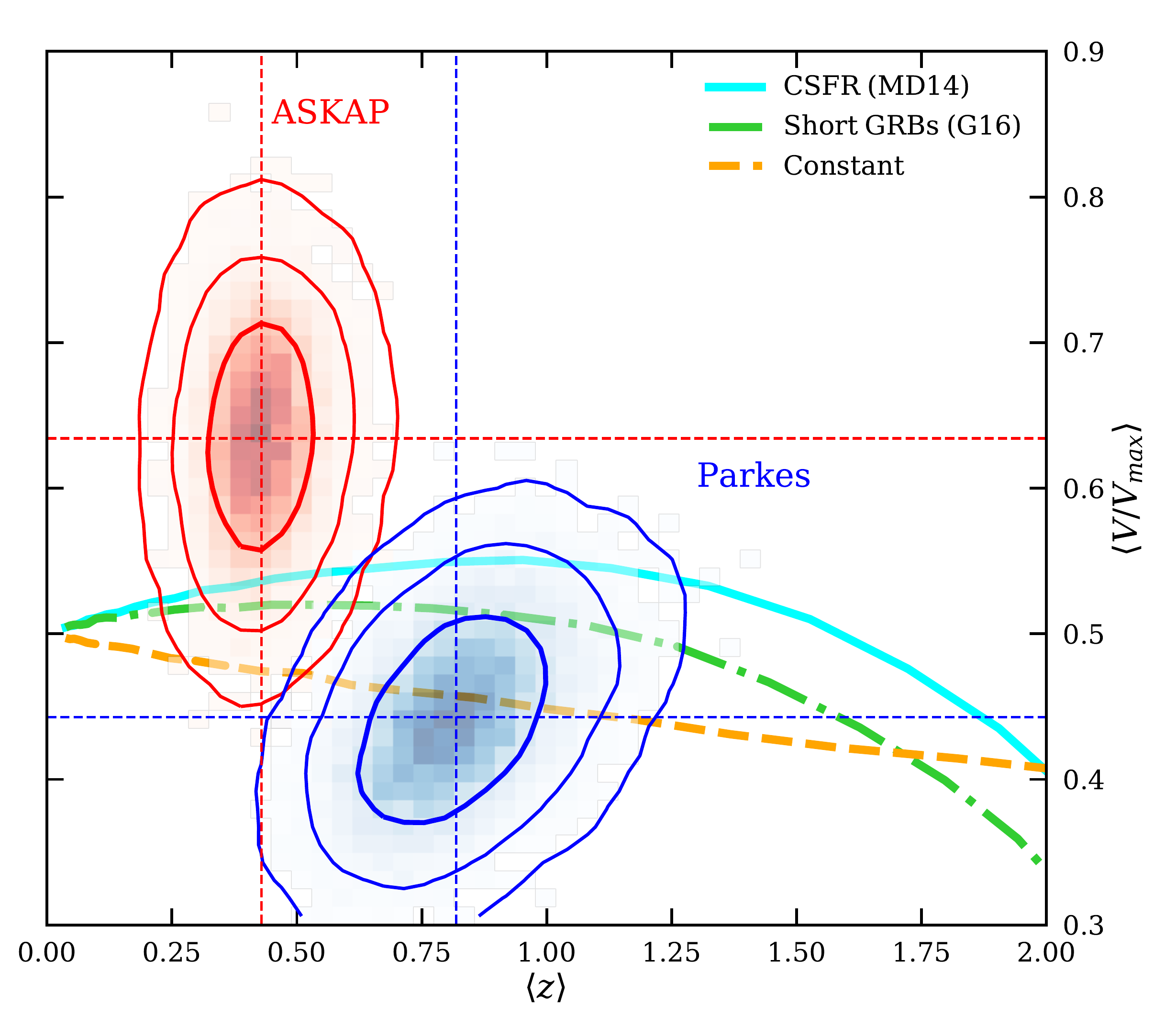}
\includegraphics[width=0.49\textwidth]{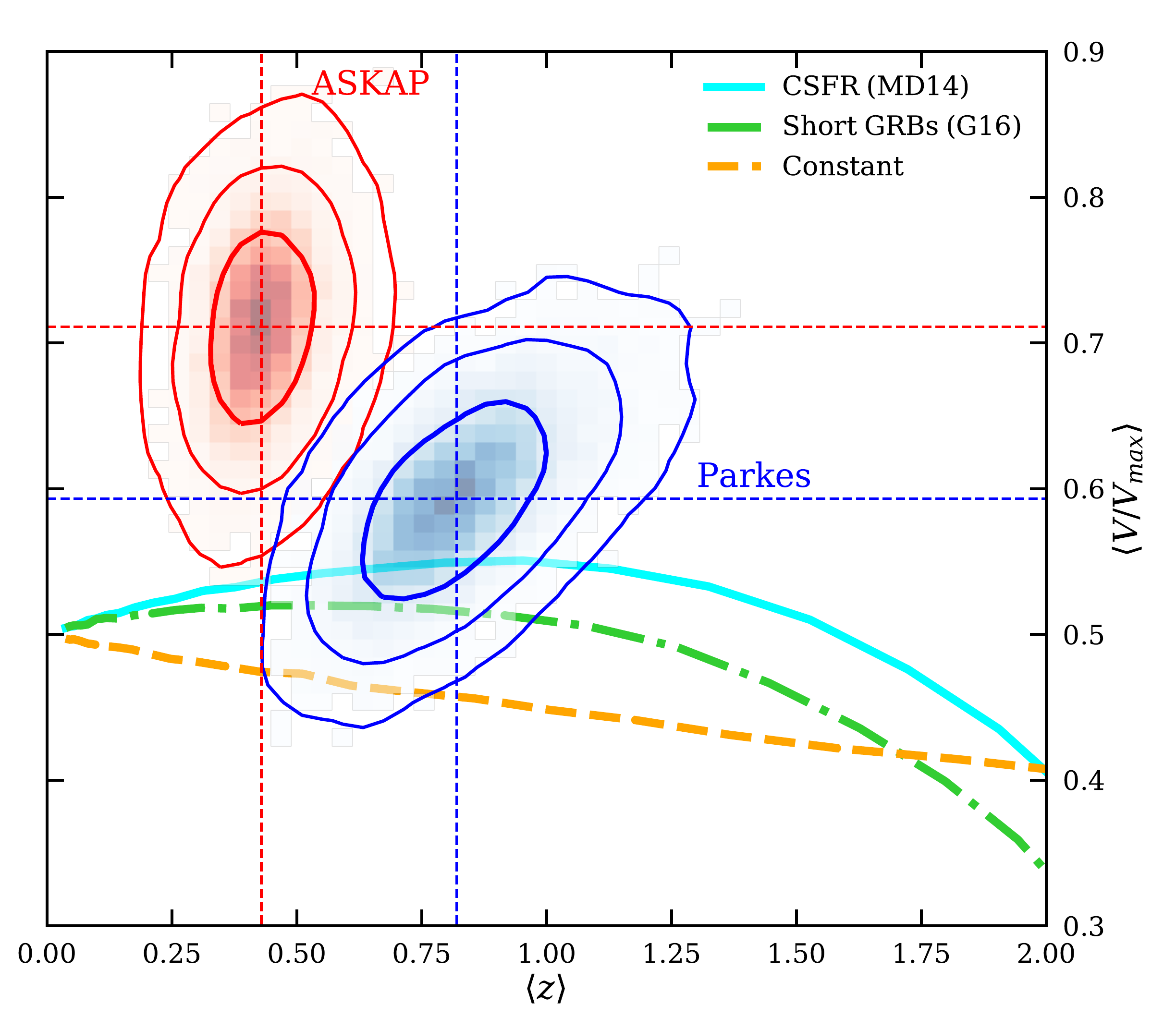}
\caption{Same as Fig.~\ref{fig:VVmVSzm} assuming different FRB spectral index $\alpha$. {\it left:} $\alpha=0$; {\it right:} $\alpha=3$}
\label{fig:VVmVSzm_alfa}
\end{figure*}
We note that the resulting curve $\langle V/V_{\rm max} \rangle$ as a function of
$\langle z \rangle$ is actually independent of the instrument 
(i.e. of the $A$ parameter). 
In fact, if the $A$ parameter is changed (namely, for different flux limits 
of the survey), at a given $z_{\rm max}$ we change the corresponding value of $E_\nu(\nu_{\rm obs})$, but the curve remains the same.

\section{Results}\label{sec:results}
The  $\langle V/V_{\rm max}\rangle$ test that we perform here includes the cosmological terms (K--correction and time dilation) arising from considering  FRBs at cosmological distances as derived from their large observed dispersion measures. The $\langle V/V_{\rm max} \rangle$ for the FRB sample (ASKAP+Parkes) and individually for the two sub--samples (ASKAP and Parkes) are shown in Table \ref{tab:V/Vm_obs} for different assumed spectral index $\alpha$ of the FRB intrinsic spectrum. 

Fig. \ref{fig:VVmVSzm} shows the $\langle V/V_{\rm max}\rangle$ values distributions, obtained from the Monte-Carlo extraction, as a function of the average redshift for the two samples for ASKAP and Parkes. 
Values obtained assuming different $\alpha$ are shown in Fig.~\ref{fig:VVmVSzm_alfa}.
Solid contours show 1-, 2- and 3-$\sigma$ level of confidence estimated via a bootstrap re-sampling of the FRB population's redshift.

The different curves show the expected $\langle V/V_{\rm max}\rangle$ as a function of mean redshift for the assumed density distributions (see \S~\ref{subsec:sim}). 
The extension of the curves before their respective turnover is due to the different redshift where the assumed density distributions peak (see also Fig.\ref{fig:FRBFR}, bottom panel).

We find the values calculated for the sub-samples differing of $\Delta\langle V/V_{\rm max}\rangle \simeq{0.12}$. We note that they probe different volumes. The Parkes $\langle V/V_{\rm max} \rangle$ is fully consistent with a cosmic stellar population (CSFR -- cyan solid line in Fig.~\ref{fig:VVmVSzm}) or its delayed version (dot--dashed green line in Fig.~\ref{fig:VVmVSzm}).
The ASKAP sample instead deviates from the CSFR scenario. 
An evidence for the difference between the ASKAP and Parkes FRB populations has also been reported by J18\cite{james18}, who find different slopes of the source count distribution for the two sub-samples.
They also report the difference to be inconsistent with the one which is expected to be due to the different volumes probed by the two surveys. 

The larger $\langle V/V_{\rm max} \rangle$ obtained for the ASKAP sample suggests a faster evolution with respect to the CSFR up to the distances currently explored by the ASKAP survey.
This large value can not be explained even considering a delayed-CSFR (green dot--dashed line in Fig.~\ref{fig:VVmVSzm})
or a different spectral index for the FRB spectrum, nor a different DM$_{\rm IGM}(z)$ relation. Overall, the ASKAP FRBs hint to a population of sources with a redshift density distribution different from those considered above (i.e. CSFR, delayed--CSFR as derived for short GRBs or constant formation rate).

\section{Discussion} \label{sec:discussion}
We have found that none of the population distributions adopted can account 
for the observed $\langle V/V_{\rm max}\rangle$.
We demonstrate here that this result is independent of the particular shape of the
energy function, as long as it does not evolve in cosmic time.
In the procedure we have adopted, each point of the curves in
Fig. \ref{fig:VVmVSzm} corresponds to a population of sources having the same
energy, calculated in such a way that its fluence corresponds to the limiting fluence
once the source is at its $z_{\rm max}$.
Smaller $\langle z\rangle$ correspond to less energetic sources.

In reality we have a distribution of energy, each corresponding to a 
different $\langle V/V_{\rm max}\rangle$, and $\langle z\rangle$.
This pairs of values, however, belong to the shown curve.
As an example, consider a specific EF, say a power law in energy 
$N(E)\propto E^{-\Gamma}$, with $\Gamma$ positive.
There will be many points at low energies, corresponding to smaller  $\langle z\rangle$
for the curve in Fig. \ref{fig:VVmVSzm}. 
Many sources at lower $\langle z\rangle$  means that the corresponding 
$\langle V/V_{\rm max}\rangle$ will be weighted more when calculating the final 
$\langle V/V_{\rm max}\rangle$. 
Still, the final value is constrained to be within the minimum and maximum values of
the curve.
As can be seen in Fig. \ref{fig:VVmVSzm} no curve can account for the 
$\langle V/V_{\rm max}\rangle$ of the ASKAP sample.

\subsection{Cosmic star formation rate} 
\label{subsec:dSFR}

Motivated by the strong hints on compact objects as sources of FRBs we consider a population with a rate density distribution  described by the cosmic star formation:
\begin{equation}
    \psi(z) = a_0 \frac{(1+z)^{a_1}}{1+\left(\frac{1+z}{1+z_p}\right)^{a_2}} \label{eq:CSFR}
\end{equation}
We adopt the parameters as reported by Madau $\&$ Dickinson (2014), $a_0 = 0.015,\, a_1=2.7, \, a_2=5.7,$ and $z_p=1.9$.
For $a_1>0$ and $a_2>1$ the formation rate peaks at $z_p$ with an increasing rate, for $z<z_p$, with slope $a_1$.
The $\left\langle V/V_{\rm max} \right\rangle$ 
as a function of $\langle z \rangle $  obtained with this function is shown in Fig. \ref{fig:VVmVSzm} (solid cyan line). 
The comparison with the ASKAP and Parkes values shows that a density evolution of FRBs following the star formation rate (Eq. \ref{eq:CSFR}) is consistent with the the Parkes  $\left\langle V/V_{\rm max} \right\rangle$
value, but is largely below the ASKAP point (which is $\sim 3\sigma$ above the expectation for a population distributed as Eq. \ref{eq:CSFR}).

\subsection{Phenomenological "FRB formation rate"} 
\label{subsec:FRBFR}

The increase with redshift ($\propto (1+z)^{p_1}$ at low $z$) of the CSFR as represented by Eq.\ref{eq:CSFR} is too shallow to account for the ASKAP point. 
In order to account for the values of $\langle V/V_{\rm max}\rangle$ of both the ASKAP and Parkes sub--samples we use the form:
\begin{equation}
{\rm FRBFR} (z) = p_0 \frac{ z^{p_1}}{1+(z/z_p)^{p_2 } }
\label{eq:FRBFR}
\end{equation}
The large $\left\langle V/V_{\rm max} \right\rangle$ of ASKAP is found 
at an average redshift significantly smaller than the redshift where the CSFR peaks
(see the bottom panel of Fig. \ref{fig:FRBFR}).
We have verified that we can reproduce both the ASKAP and the Parkes 
$\left\langle V/V_{\rm max} \right\rangle$ values using
$p_0=1.$
$p_1=3.7$, 
$p_2=4.8$ 
and $z_{p} = 0.6$. We stress that this does not correspond to a formal fit and other possible functional forms could well be consistent with the two points. Proper model selection and parameter fitting is out of the scope of the current work. Here we want to find an empirical density distribution which we can a-posteriori compare with the cosmic star formation rate. 

The model described above is shown by the solid red line in Fig. \ref{fig:FRBFR} (top panel). 
For comparison, also the model curve obtained from the CSFR described in the previous section is shown (dotted cyan line).

\subsection{Delayed Cosmic Star Formation Rate}\label{subsec:delay}

\begin{figure}
    \centering
    \hskip -0.4 cm
    \includegraphics[width=0.51\textwidth]{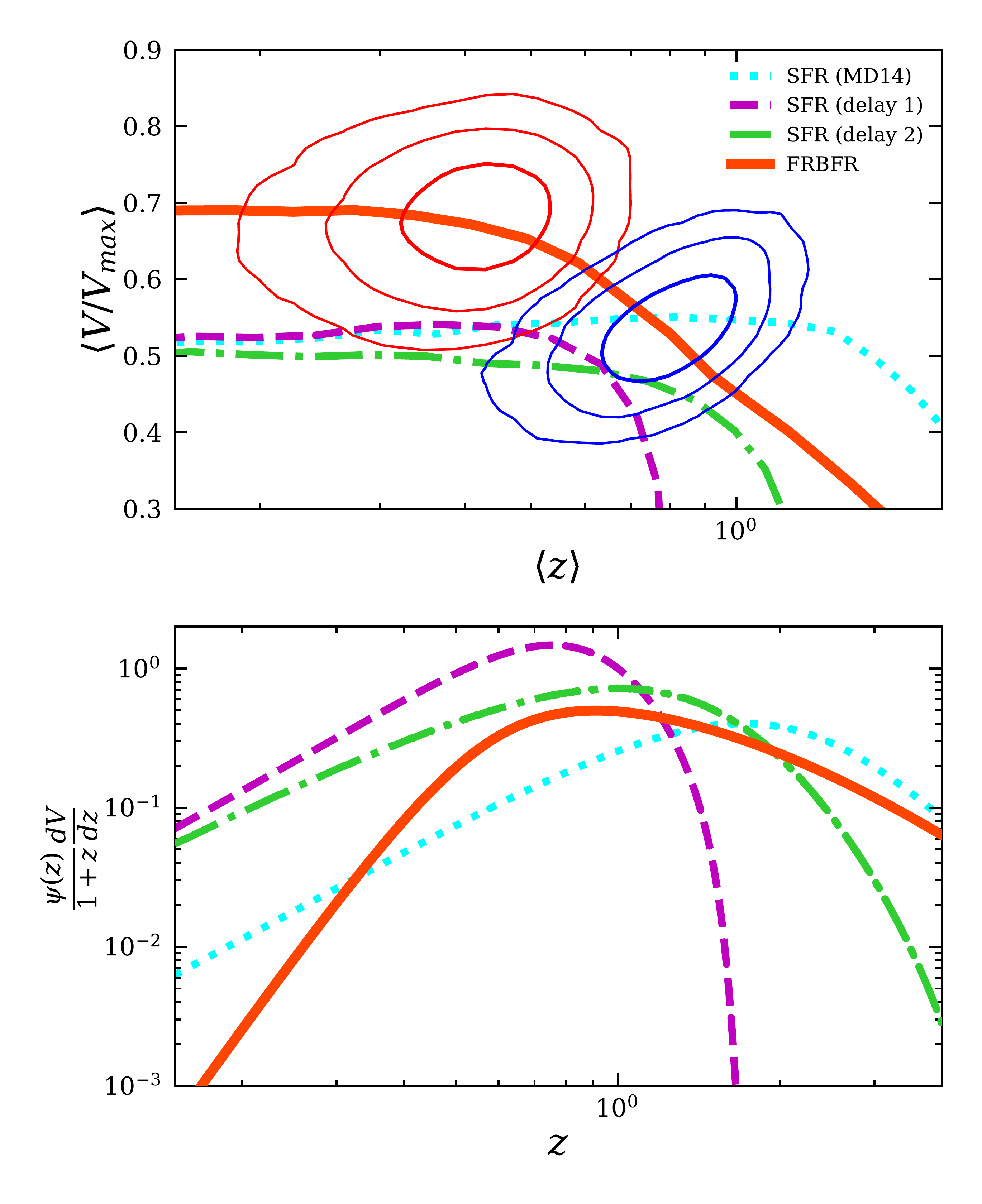}
    \caption{Simulated populations obtained from different redshift density distributions $\psi_k(z)$.
    Upper panel: $\left\langle V/V_{\rm max} \right\rangle$ as function of the average observable redshift. Contours show the ASKAP and Parkes confidence levels assuming $\alpha=1.6$ (see Fig.$~\ref{fig:VVmVSzm}$). 
    Bottom panel: normalised source rate as function of redshift. Distributions are normalised to their integral. }
    \label{fig:FRBFR}
\end{figure}

One possibility, e.g. motivated by the similar case of short GRBs, is that the progenitors of FRBs produce the radio flashes at some advanced stage of their evolution. This would introduce a delayed density distribution which in the simplest case can be modelled starting from the CSFR as (see G16)
\begin{equation}
\Psi(z) = \int_z^\infty \psi(z')P(t(z)-t(z')) \frac{dt}{dz'}dz'
\end{equation}
For the delay function we consider (1) a normal PDF centred on 4~Gyr with a dispersion $\sigma=0.1$~Gyr, and (2) a power-law with slope $\propto \tau^{-1}$ with a minimum delay equal to $1$ Gyr. 
The latter is what is expected for the merging of compact objects (Greggio 2005\cite{greggio05}; Belczynski et al. 2006\cite{belczynski06}; Mennekens et al. 2010\cite{mennekens10};  Ruiter et al. 2011\cite{ruiter11};  Mennekens $\&$ Vanbeveren 2016\cite{mennekens16}, Cao et al. 2018\cite{cao18}). 
The $\left\langle V/V_{\rm max} \right\rangle$ curves obtained with these two models are shown in Fig.~\ref{fig:FRBFR} (top panel) by the dashed (magenta) and dot--dashed (green) lines, respectively. 
The resulting source rate are plotted in Fig.~\ref{fig:FRBFR} (bottom panel) in purple and green colour respectively. 
Although they peak at later times by construction, none of the delay models look to be consistent with the ASKAP data. 
In fact the effect of the delay is just to move the maximum value attainable for the $\left\langle V/V_{\rm max} \right\rangle$ to later times (i.e. smaller redshifts), but does not increase enough to become consistent with the ASKAP value (cfr Fig~\ref{fig:FRBFR}, upper panel).

\section{Conclusions}

We have performed the $\langle V/V_{\rm max}\rangle$ test for two different FRB samples,
assuming that their distances are cosmological.
Having characterised each sample through its average redshift, 
we have verified that the value of $\langle V/V_{\rm max}\rangle$ for the Parkes
sample is similar to the one derived  assuming that FRBs follow the star formation 
rate.
This would indicate that FRBs can be associated to compact stars, 
such as neutron stars.
Instead, for the ASKAP sample, sampling a closer volume but slightly 
more energetic FRBs, the value of $\langle V/V_{\rm max} \rangle$ is larger 
than the one derived with a population following the star formation rate.
This result depends negligibly on the assumed FRB spectral index and 
has been determined accounting 
for the different sources of uncertainty on the estimate of the redshift of FRBs 
taking on a conservative approach.

At face value, this suggests that the progenitors of FRBs evolve in 
cosmic time faster than the star formation, and reach their maximum 
at redshift between the average redshift of the ASKAP and Parkes samples.
Such fast evolution at relatively low redshifts can be due to 
a density evolution faster than the star formation rate, or to 
a luminosity (and energy) evolution superposed to a standard star 
formation rate.
This is intriguing, because it would suggest a very peculiar 
population for the FRB progenitors, but at this stage we are not able to 
disentangle between these possibilities.
A possibility is that, similarly to short Gamma Ray Bursts, there is a 
delay between the formation of the stars producing the FRB phenomenon and 
the FRB event. 
However, we verified that this cannot produce the steep rise 
(as a function of redshift)
of event rate up to $z\sim$ 0.3--0.4 needed to account for the observed
ASKAP $\langle V/V_{\rm max}\rangle$.
The phenomenological ``FRB formation rate" we have found, that can fit 
both the ASKAP and Parkes $\langle V/V_{\rm max}\rangle$ cannot be interpreted in
a simple way on the basis of known population of sources.
While the investigations of possibilities is demanded to a future work,
we stress the need of having a survey exploring the same (large) sky area
of ASKAP, but with a fluence limit comparable to Parkes.
This will show how the FRB population evolves in time in the interesting
0.3--2 redshift range.

\section*{Acknowledgments}
NL aknowledges financial support from the Horizon 2020 program under the ERC Starting Grant "MAGCOW", no. 714196. 
We aknowledge the PRIN-INAF 2017 "Towards the SKA and CTA era: discovery, localisation, and physics of transient sources".

\begin{table*}
\begin{tabular}{l|llllllllll}
\hline
  FRB name &$gl$ &$gb$ &$DM_{\rm obs}$ &$S/N$ &$S/N_{\rm lim}$ &Fluence &$w_{\rm obs}$ &Flux &$z\,\dagger$  &Ref. \\
    & deg & deg & pc/cm$^3$ &  &  &Jy ms &ms &Jy   & \\
\hline
~   &  & & & ASKAP \\
\hline
 180525	&349.0	 &50.7 	    &388.1	&27.4   &9.5 & 300  & 3.8   &  78.9    & 0.174$_{-0.123}^{+0.391}$    &Ma18   \\
 180324	&245.2	 &--20.5    &431	&9.8    &9.5 & 71   & 4.3   &  16.51     & 0.172$_{-0.121}^{+0.392}$    &Ma18   \\
 180315	&13.2	 &--20.9	&479	&10.4   &9.5 & 56   & 2.4   &  23.34    & 0.215$_{-0.142}^{+0.396}$    &Ma18   \\
 180212	&338.3	 &50.0	    &167.5	&18.3   &9.5 & 96   & 1.81  &  53.04    & 0.022$_{-0.020}^{+0.321}$    &Sh18   \\
 180131	&0.6	 &--50.7	&657.7	&13.8	&9.5 & 100  & 4.5   &  22.22     & 0.420$_{-0.210}^{+0.403}$    &Sh18   \\
 180130	&5.9	 &--51.8	&343.5	&10.2	&9.5 & 95   & 4.1   &  23.17     & 0.132$_{-0.099}^{+0.390}$    &Sh18   \\
 180128.2 &327.8 &--48.6	&495.9	&9.6	&9.5 & 66   & 2.3   &  28.7     & 0.270$_{-0.164}^{+0.399}$    &Sh18   \\
 180128.0 &326.7 &52.2	    &441.4	&12.4	&9.5 & 51   & 2.9   &  17.6     & 0.221$_{-0.144}^{+0.397}$    &Sh18   \\
 180119	&199.5	 &--50.4	&402.7	&15.9	&9.5 & 110  & 2.7   &  40.74     & 0.175$_{-0.123}^{+0.394}$    &Sh18   \\
 180110 &7.8	 &--51.9	&715.7	&35.6	&9.5 & 420  & 3.2   &  131.25     & 0.472$_{-0.222}^{+0.406}$    &Sh18   \\
 171216	&273.9	 &--48.4	&203.1	&10.3  	&9.5   & 40   & 1.9   &  21.06    & 0.034$_{-0.030}^{+0.345}$   &Sh18    \\
 171213	&200.6	 &--48.3	&158.6	&25.1	&9.5 & 133  & 1.5   &  88.67    & 0.013$_{-0.012}^{+0.308}$    &Sh18   \\
 171116	&205.0	 &--49.8	&618.5	&11.8	&9.5 & 63   & 3.2   &  19.69     & 0.372$_{-0.198}^{+0.404}$    &Sh18   \\
 171020	&29.3	 &--51.3	&114.1	&19.5	&9.5 & 200  & 3.2   &  62.5      & 0.006$_{-0.006}^{+0.282}$    &Sh18   \\
 171019	&52.5	 &--49.3	&460.8	&23.4	&9.5 & 219  & 5.4   &  40.56     & 0.245$_{-0.155}^{+0.391}$    &Sh18   \\
 171004	&282.2	 &48.9  	&304	&10.9	&9.5 & 44   & 2     &  22.       & 0.095$_{-0.075}^{+0.381}$    &Sh18   \\
 171003	&283.4	 &46.3	    &463.2	&13.8	&9.5 & 81   & 2     &  40.5       & 0.232$_{-0.150}^{+0.398}$    &Sh18   \\
 170906	&34.2	 &--49.5	&390.3	&17	&9.5 & 74   & 2.5   &  29.6      & 0.173$_{-0.123}^{+0.395}$     &Sh18   \\ 
 170712	&329.3	 &--51.6	&312.79	&12.7	&9.5 & 53   & 1.4   &  37.86 	 & 0.109$_{-0.084}^{+0.383}$  &Sh18   \\
 170707	&269.1	 &--50.5	&235.2	&9.5	&9.5 & 52   & 3.5   &  14.86 	 & 0.053$_{-0.045}^{+0.361}$   &Sh18   \\
 170428	&359.2	 &--49.9	&991.7	&10.5	&9.5 & 34   & 4.4   &  7.73 	 & 0.716$_{-0.263}^{+0.416}$   &Sh18   \\
 170416	&337.6	 &--50	    &523.2	&13 	&9.5 & 97   & 5     &  14.4  	 & 0.293$_{-0.173}^{+0.402}$   &Sh18   \\
 170107	&266.0	 &51.4	    &609.5	&16	   &9.5  & 58   & 2.4   &  24.17 	 & 0.376$_{-0.198}^{+0.402}$  &Ba17   \\
\hline
~  & & & & Parkes \\
\hline
 160102	&18.9	 &--60.8    &2596.1 	&16	&10 &  1.8    &	3.4   & 1.06 	&  2.097$_{-0.381}^{+0.473}$   &Bh18  \\
 151230	&239.0	 &34.8	    &960.4	    &17	&10 &  1.9    &	4.4   & 0.86 	&  0.682$_{-0.257}^{+0.412}$   &Bh18     \\
 151206	&32.6	 &--8.5	    &1909.8	    &10	&10 &  0.9    &	3     & 0.6  	&  1.395$_{-0.332}^{+0.445}$   &Bh18     \\
 150610	&278.0	 &16.5	    &1593.9	    &18	&10 &   1.3    &  2       & 0.53   &   1.155$_{-0.312}^{+0.437}$   &Bh18   \\
 150418	&232.7 &--3.2	&776.2	    &39	&10&   1.76   &  0.8     & 2.2	 &   0.247$_{-0.157}^{+0.405}$   &Ke16   \\
 150215	&24.7 &5.3	&1105.6	    &19	&10 &   2.02   &  2.88    & 0.7	 &   0.582$_{-0.242}^{+0.411}$   &Pe17   \\
 140514	&50.8 &--54.6	&562.7	    &16	&10&   1.32   &  2.82    & 0.47  &   0.331$_{-0.186}^{+0.404}$   &Pe15   \\
 131104	&260.6	 &--21.9	&779	    &34	&10 &   2.75   &  2.37    & 1.16  &   0.352$_{-0.192}^{+0.403}$   &Sh16   \\
 130729	&324.8 &54.7	&861	    &14	&10&   3.43   &  15.61   & 0.22  &   0.602$_{-0.246}^{+0.411}$   &Ch15   \\
 130628	&226.0 &30.7	&469.88     &29	&10&   1.22   &  0.64    & 1.91  &   0.230$_{-0.149}^{+0.396}$   &Ch15   \\
 130626	&7.5 &27.4	&952.4	    &21	&10&   1.47   &  1.98    & 0.75  &   0.648$_{-0.253}^{+0.413}$  &Ch15   \\
 121002	&308.2	 &--26.3	&1629.18    &16	&10&   2.34   &  5.44    & 0.43  &   1.240$_{-0.319}^{+0.439}$   &Ch15   \\
 120127	&49.3 &--66.2	&553.3	    &13	&9 &   0.75   &  1.21    & 0.62  &   0.331$_{-0.186}^{+0.399}$   &Th13   \\
 110703	&81.0 &--59.0	&1103.6	    &17	&9 &   1.75   &  4.3     & 0.41  &   0.817$_{-0.276}^{+0.422}$   &Th13   \\
 110626	&355.9 &--41.8	&723	    &12	&9 &   0.89   &  1.41    & 0.63  &   0.471$_{-0.222}^{+0.407}$   &Th13   \\
 110220	&50.8	 &--54.8	&944.38	    &54	&9 &  7.31   &  6.59    & 1.11  &   0.674$_{-0.257}^{+0.417}$    &Th13   \\
 110214   &290.7   &--66.6    &168.9    &13 	&5 &  51.3  &	1.9    &  27.	   &  0.022$_{-0.020}^{+0.329}$    &Pe18 \\
 090625	&226.4 &--60.0	&899.55	    &30	&10&  2.19  &	1.92   &  1.14	   &  0.634$_{-0.251}^{+0.415}$     &Ch15   \\
 010621	&25.4	&	-4.0	&	745	& 18	& 8	& 4.24  &	8.     &  0.53     &  0.228$_{-0.148}^{+0.399}$	& Pe16	\\
 010125	&356.6 &--20.0	&790	    &25 &7 &  5.72  &	10.6   & 0.54 &  0.495$_{-0.226}^{+0.406}$     &Ke12 \\
\hline 
\end{tabular}
\caption{Observational and instrumental parameters of FRBs in our sample. In the last column we report the references for the S/N$_{\rm lim}$:
Ma18: Macquart et al. (2018);
Sh18: Shannon et al. (2018),
Ba17: Bannister et al. 2017;
Bh18: Bhandari et al. (2018);
Ke17: Keane et al. (2016);
Pe17: Petroff et al. (2017);
Pe15: Petroff et al. (2015);
Sh16: Shand et al. (2016);
Ch15: Champion et al. (2015);
Sc16: Sholtz et al. (2016);
Th13: Thornton et al. (2013);
Pe18: Petroff et al. (2018);
ke12: Keane et al. (2012).
Parkes FRBs parameters are taken from Petroff et al. (2016) when available, or from the reference in the last column.
$\dagger$: Redshifts $z$ are calculated from DMs as in $\S 3$.
}
\label{tab:sample_obs}
\end{table*}


\section*{\appendixname{ }} 
\subsection*{$\langle V/V_{\rm max}\rangle$ value estimation}
\label{subsec:derivation}

The $V/V_{\rm max}$ computed in this work (Eq.~\ref{eq:D/Dm}), which accounts for the cosmological k-correction and proper rate of transient FRBs, is derived as follows. 
We consider the relation:
\begin{equation}
{\cal F}(\nu_{\rm obs}) =  \frac{(1+z)E_{\nu}(\nu_e)}{4\pi D_L^2(z) } \frac{d\nu_e}{d\nu_{\rm obs}} = \frac{E_{\nu}(\nu_e)}{4\pi D^2(z) }  \label{eq:F(nu_obs)}
\end{equation}
where we use the relation between luminosity distance and proper distance: $D_L = (1+z)D$. Analogously:
\begin{equation}
{\cal F}_{\rm lim}(\nu_{\rm obs}) = \frac{E_{\nu}(\nu'_e)}{4\pi D_{\rm max}^2(z_{\rm max})} \cdot \label{eq:F(nu_obs)_lim}
\end{equation}
 where $E_\nu$ is the intrinsic energy density, $\nu_e=\nu_{\rm obs}(1+z)$ and $\nu'_e=\nu_{\rm obs}(1+z_{\rm max})$ are the emitted frequencies at proper distances $D(z)$ and $D_{\rm max}(z_{\rm max})$, respectively.
 
We assume that FRBs have a power-law energy spectrum spectral slope $\alpha$, at least in a frequency range comparable to that probed by the observer frame frequency $\nu_{\rm obs}\simeq 1.3$~GHz. Therefore, we can write:
\begin{eqnarray}
{dE} & \propto&\nu^{-\alpha}  d\nu  \nonumber \\
E_{\nu}(\nu_e)&=& E_{\nu}(\nu'_e) \cdot \left(\frac{\nu_e}{\nu'_e}\right)^{-\alpha} \nonumber \\
&=& E_{\nu}(\nu'_e) \cdot \left(\frac{\nu_{\rm obs}\,(1+z)}{\nu_{\rm obs}\,(1+z_{\rm max})}\right)^{-\alpha}\nonumber \\ 
&=& E_{\nu}(\nu'_e)\cdot \left( \frac{1+z}{1+z_{\rm max}} \right)^{-\alpha}
\label{eq:lum_rel}
\end{eqnarray}
Combining  eq.\ref{eq:F(nu_obs)},~\ref{eq:F(nu_obs)_lim} and~\ref{eq:lum_rel} we obtain: 
\begin{equation}
 \frac{D(z)}{D_{\rm max}(z_{\rm max})} = \left[ \frac{F_{\rm lim}(\nu_{\rm obs})}{F(\nu_{\rm obs})} \left( \frac{1+z}{1+z_{\rm max}}\right)^{-\alpha}\right]^{1/2}. \label{eq:D_F_rel}
\end{equation}
From the antenna equation, the signal-to-noise ratio of an event of duration $w_{\rm obs}$ is defined as: 
\begin{equation}
    S/N = \frac{G\, S(\nu) \sqrt{N_P \Delta\nu \, w_{\rm obs}} }{\eta T_{\rm sys}}
\end{equation}
where G is the antenna gain, $N_P$ is the number of polarizations, $\eta$ is the efficiency and $T_{\rm sys}$ is the antenna temperature. The S/N can also be described in terms of the fluence ${\cal F}(\nu) = S(\nu)\, w_{\rm obs}$ as:
\begin{equation}
    S/N = \frac{{\cal F}(\nu)}{\sqrt{w_{\rm obs}}}\frac{G \sqrt{N_P \Delta\nu } }{\eta T_{\rm sys} }. \label{eq:S/N_def}
\end{equation} 
The limiting threshold over which an FRB is detected can also be described in terms of S/N: 
\begin{equation}
    S/N_{\rm lim} = \frac{{\cal F}_{\rm lim}(\nu)}{\sqrt{w_{\rm obs}'}}\frac{G \sqrt{N_P \Delta\nu } }{\eta T_{\rm sys} }. \label{eq:S/Nlim_def}
\end{equation}
where $w_{\rm obs}'= w_{\rm rest}(1+z_{\rm max})$; all the parameters related to the observing conditions/setup ($G,\, N_P,\, \Delta \nu,\, \eta $ and $T_{\rm sys}$) are the same when we want to look for the maximum distance at which the FRB could be observed. 
However, the observed duration of the transient event changes with redshift. 
We can thus relate the ratios of observed and threshold values with a function of their redshifts:
\begin{equation}
  \frac{{\cal F}(\nu)}{{\cal F}_{\rm lim}(\nu)} =  \frac{S/N}{S/N_{\rm lim}}  \left(\frac{1+z}{1+z_{\rm max}}\right)^{1/2} \label{eq:F_S/N_rel}  
\end{equation} 
Through Eq.~\ref{eq:F_S/N_rel} we can recast Eq.~\ref{eq:D_F_rel} in terms of the $S/N$ with respect to $S/N_{\rm lim}$ \begin{equation}
  \frac{D(z)}{D_{\rm max}(z_{\rm max})} =  \left[ \frac{S/N_{\rm lim}}{S/N}  \left(\frac{1+z}{1+z_{\rm max}}\right)^{-\frac{1}{2}-\alpha}\right]^\frac{1}{2} \label{eq:D/Dm_appendix}
\end{equation}
retrieving Eq.~\ref{eq:D/Dm}. Equivalently one can write it as 
\begin{equation}
  \frac{S/N}{S/N_{\rm lim}} \, D^2(z) \, (1+z)^{\frac{1}{2}+\alpha} =   D_{\rm max}^2(z)\,  (1+z_{\rm max})^{\frac{1}{2}+\alpha} \equiv \Upsilon(z_{\rm max}) \label{eq:ups_def}
\end{equation}

We put in the right-hand side (RHS) of eq.~\ref{eq:ups_def} all the terms depending on $z_{\rm max}$. We define the function $\Upsilon(z_{\rm max})$ as the RHS. $\Upsilon(z_{\rm max})$ can be evaluated using the left-hand side (LHS) of the same equation, i.e. combining the observed, instrumental and cosmological information.
If the functions $D_{\rm max}(z)$ and $\Upsilon(z)$ are invertible one can in principle solve the above equation for $z_{\rm max}$. We note that only for $\alpha>-1/2$ this function is monotonic and invertible. Under this assumption
\begin{equation}
z_{\rm max} = \Upsilon^{-1} \left[\frac{S/N}{S/N_{\rm lim}} \, D^2(z) \, (1+z)^{\frac{1}{2}+\alpha} \right]. \label{eq:z_max}
\end{equation}
The function $\Upsilon$ depends on cosmology through the definition of $D(z)$
\begin{equation}
D(z) = \int_0^z \frac{c \,dz'}{H_0\sqrt{\Omega_{\rm max}(1+z')^3+\Omega_k (1+z')^2 + \Omega_{\Lambda}}}\label{eq:D_max}
\end{equation}
and from the spectral index $\alpha$ of the intrinsic FRB specific luminosity. 

The estimates of the spectral index is given considering the 23 ASKAP burst signals (Shannon et al., 2018) detected with narrow--band (336~MHz) centred at 1.32 GHz showing intense fine--scale features (Macquart et al., 2018\cite{macquart18}). 
They calculate a mean spectral index $\alpha=1.6^{+0.3}_{-0.2}$. 
This represents the current best (and only) estimate of the spectral index for the non-repeating FRBs, as far as any broad-band information will be given.

The analytic form of Eq. \ref{eq:z_max} is not straightforward to obtain so we solved it numerically.  

\subsection*{$S/N$ and $S/N_{\rm lim}$}

The fact itself that in Eq. \ref{eq:z_max} the ratio between a threshold and the corresponding observed quantity are present (instead of just the observed one) removes the dependency of the test from any survey parameter, since they would play the same role in the definition of both terms in the ratio.
The fact that the ratio between observed and threshold S/N is 
independent of the survey area and time coverage or any other observation parameter 
has been previously proven by J18\cite{james18} in a rigorous way. 
Our derivation can be helpful in giving an intuitive and straightforward proof of the fact.

The $(S/N)/(S/N)_{\rm lim}$ approach has also the advantage that any other quantity which varies linearly with the fluence ${\cal F}(\nu)$ can be used in its place in Eq. \ref{eq:z_max} {${}^{\cite{oppermann16}}$}. 
The smartest choice is to use the signal-to-noise ratio ($S/N$), which is the ratio between the amplitude of the time-integrated FRB signal (after the noise level being normalized to 0) and the standard deviation of the noise in the continuum (Petroff et al. 2016).

We choose to use $S/N$ as $S/N_{\rm lim}$ is the quantity which is actually defined in a survey in order to claim a detection, rather than the flux density $S_{\nu}$ or the fluence ${\cal F}_{\nu}$.
Moreover, the way the $S/N$ is defined is in principle independent from the pulse broadening effect whenever a signal is detected. 
In fact, highly--broadened signals could not be recognized as FRB candidates by searching pipelines, but a non-detection does not affect the completeness of the sample in a $\left\langle V/V_{\rm max} \right\rangle$ test. 

\end{document}